\documentclass[12pt,english,aps,prl,twocolumn,showpacs,superscriptaddress,groupedaddress,footinbib,reprint,noshowpacs]{revtex4-1}
\usepackage{graphicx}
\usepackage{amsmath}
\usepackage{comment}
\usepackage{amssymb}

\usepackage{subfigure}

\newcommand{\Ffluct}{F_{\mathrm{fluct}}}
\newcommand{\kmax}{k_{\mathrm{max}}}

\begin{document}

\title{Fluctuation spectra and force generation in non-equilibrium systems}

\author{Alpha A. Lee}
\email{alphalee@g.harvard.edu}
\affiliation{School of Engineering and Applied Sciences and Kavli Institute of Bionano Science and Technology, Harvard University, Cambridge, MA 02138, USA}

\author{Dominic Vella}
\email{dominic.vella@maths.ox.ac.uk}
\affiliation{Mathematical Institute, Andrew Wiles Building, University of Oxford, Woodstock Road, Oxford OX2 6GG, UK}

\author{John S. Wettlaufer}
\email{john.wettlaufer@yale.edu}
\affiliation{Mathematical Institute, Andrew Wiles Building, University of Oxford, Woodstock Road, Oxford OX2 6GG, UK}
\affiliation{Yale University, New Haven, USA}
\affiliation{Nordita, Royal Institute of Technology and Stockholm University, SE-10691 Stockholm, Sweden}

\begin{abstract}
Many biological systems are appropriately viewed as passive inclusions immersed in an active bath: from proteins on active membranes to microscopic swimmers confined by boundaries. The non-equilibrium forces exerted by the active bath on the inclusions or boundaries often regulate function, and such forces may also be exploited in artificial active materials. Nonetheless, the general phenomenology of these active forces remains elusive. We show that the fluctuation spectrum of the active medium, the partitioning of energy as a function of wavenumber, controls the phenomenology of force generation. We find that for a narrow, unimodal spectrum, the force exerted by a non-equilibrium system on two embedded walls depends
on the width and the position of the peak in the fluctuation spectrum, and oscillates between repulsion and attraction as a function of wall separation. We examine two apparently disparate examples: the Maritime Casimir effect and recent simulations of active Brownian particles. A key implication of our work is that important non-equilibrium interactions are encoded within the fluctuation spectrum. In this sense the noise becomes the signal.
\end{abstract}

\makeatother
\maketitle

Force generation between passive inclusions in active, non-equilibrium systems underpins many phenomena in nature. Bioinspired examples in which such interactions might arise range from proteins on active membranes \cite{manneville1999activity,gov2004membrane} to swimmers confined by a soft boundary \cite{Spellings2015,paoluzzi2016,takatori2016acoustic}. On the large scale, such systems feature interactions between objects in a turbulent flow and ships on a stormy sea \cite{boersma1996maritime}. A fundamental physical question arising is whether there is a convenient physical framework that could describe force generation in the wide variety of out-of-equilibrum systems across different lengthscales? 

The salient challenge is that, unlike an equilibrium system, the continuous input of energy places convenient and general statistical concepts, underlying the partition function and the free energy, on more tenuous ground. For example, theories and simulations of active Brownian particles show that self-propulsion induces complex phase behavior qualitatively different from the passive analogue \cite{fily2012athermal,cates2012diffusive,redner2013structure,stenhammar2013continuum,buttinoni2013dynamical,cates2015motility}, and non-trivial behavior such as flocking and swarming is realizable in a non-equilibrium system \cite{toner1995}. Therefore, many studies focus on the microscopic physics of a particular active system to compute the force exerted on the embedded inclusions \cite[e.g.,][]{angelani2011effective,ray2014casimir,parra2014casimir,mallory2014anomalous,harder2014role,leite2016depletion,bechinger2016active}.  

In this paper we show that the force generated by an active system on passive objects is determined by the partition of energy in the active system, given mathematically by the wavenumber dependence of energy fluctuations within it.  A key prediction is that, if the energy fluctuation spectrum is non-monotonic, the force can oscillate between attraction and repulsion as a function of the separation between objects. By making simple approximations of a narrow, uni-modal spectrum, we extract scaling properties of the fluctuation-induced force that compare well with recent simulations of the force between solid plates in a bath of self-propelling Brownian particles \cite{ni2015tunable}.

\section*{Fluctuation Spectrum and Fluctuation-Induced Force}

We begin with the question: How can we distinguish a suspension of pollen grains at thermal equilibrium from a suspension of active microswimmers? On the one hand, it has been shown that 
the breakdown of the fluctuation dissipation relation may be directly probed \cite{mizuno2007nonequilibrium,turlier2016equilibrium,battle2016broken}, and, further, that novel fluctuation modes emerge out of equilibrium \cite{rodriguez2015direct}.  On the other hand, an alternative way to characterize the system is via the wavenumber-dependent energy fluctuation spectrum.
  A natural means of monitoring the fluctuation spectrum (the spectrum of noise due to random forces in the particles' dynamics) uses dynamic light scattering \cite{chu1974laser}. A general feature of the macroscopic view of physical systems is that fluctuations are intrinsic due to statistical averaging over microscopic degrees of freedom. The magnitude of this intrinsic noise can in general be a function of the frequency and wavenumber --- this fluctuation spectrum is one key signature of a particular physical system. 

Although the fluctuation spectrum can be derived from microscopic kinetic processes, here we are interested in showing that the general properties of such spectra can provide a framework for understanding nonequilibrium behavior. Equilibrium thermal fluctuations, such as that for a Brownian suspension or Johnson--Nyquist noise \cite{nyquist1928thermal}, are usually associated with white noise corresponding to equipartition of energy between different modes. The key point here is that non-equilibrium processes have the potential to generate a nontrivial (even non-monotonic) fluctuation spectrum by continuously injecting energy into particular modes of an otherwise homogenous medium. In the example of microswimmers, they create ``active turbulence'' by pumping energy preferentially into certain lengthscales of a homogeneous isotropic fluid \cite{wensink2012meso}.

The relation between fluctuation spectra and disjoining force may be examined by generalizing the classic calculation of Casimir \cite{casimir1948attraction}. We consider an effectively one dimensional system of two infinite, parallel plates separated by a distance $L$ and immersed in a non-equilibrium medium. We assume that the fluctuations are manifested as waves and  impart a radiative stress. We define the fluctuation spectrum as
\begin{equation}
G(k) \equiv \frac{\mathrm{d} E(k)}{ \mathrm{d}k}, 
\label{fluct_spec}
\end{equation} 
where $E(k)$ is the energy density of modes with wavenumber $k$.  Whence the radiation force per unit plate area, $\delta F $, due to waves with wavenumber between $k$ and $k+\delta k$ (where $k = |\mathbf{k}|$ is the magnitude of the wavevector), and with angle of incidence between $\theta$ and $\theta+\delta\theta$, is 
\begin{equation}
\delta F =  G(k) \delta k \cos^2 \theta \frac{\delta \theta}{2\pi}. 
\label{intensity_force}
\end{equation} 
One factor of $\mathrm{cosine}$ in Eq. (\ref{intensity_force}) is due to projecting the momentum in the horizontal direction, the other factor of $\mathrm{cosine}$ is due to momentum being spread over an area larger than the cross sectional length of the wave, and the factor of $2\pi$ accounts for the force per unit angle (see e.g.~\cite{landau1987fluid} for a more detailed derivation of Eq (\ref{intensity_force})). For isotropic fluctuations, we can consider $\delta \theta$ as an infinitesimal quantity and, upon integrating from $\theta = -\pi/2$ to $\pi/2$, we arrive at 
\begin{equation}
\delta F =  \frac{1}{4} G(k) \delta k. 
\end{equation}  

Outside of the plates, any wavenumber is permitted and so
\begin{equation}
F_{\mathrm{out}} = \frac{1}{4}  \int_{0}^{\infty}G(k) \mathrm{d}k. 
\end{equation}
However,  the waves traveling perpendicular to and between the plates are restricted to take only integer multiples of $\Delta k = \pi/L$, because the waves are reflected by each plate. The force imparted by the waves to the inner surface of the plates is then
\begin{equation}
F_{\mathrm{in}} = \frac{1}{4} \sum_{m=1}^{\infty}  G (m \Delta k) \; \Delta k
\label{eqn:Fin}
\end{equation}
in one dimension. Thus, the \emph{net} disjoining force for a one dimensional system is given by 
\begin{equation}
\Ffluct= F_{\mathrm{in}} - F_{\mathrm{out}} = \frac{1}{4} \sum_{m=1}^{\infty}  G(m \Delta k) \; \Delta k - \frac{1}{4}\int_{0}^{\infty} G(k) \; \mathrm{d}k. 
\label{central_res}
\end{equation}
Note that $\Ffluct \lessgtr 0$ for all plate separations $L$ if the derivative $G'(k)\lessgtr 0$ for all $k$:  if a non-monotonic force is observed, it necessarily implies a non-monotonic spectrum.  Furthermore, in higher dimensions the continuous modes need to be integrated to compute the force between the plates. 

Clearly, the fluctuation spectrum $G(k)$ is the crucial quantity in our framework, and can, in principle, be calculated for different systems. We note that previous theoretical approaches have mostly focused on the stress tensor \cite{dean2010out}. For example, the effect of shaking protocols on force generation have been investigated theoretically for soft \cite{bartolo2003effective} and granular \cite{cattuto2006fluctuation} media. More generally, non-equilibrium Casimir forces have been computed for reaction-diffusion models with a broken fluctuation-dissipation relation \cite{brito2007generalized,rodriguez2011dynamical}, and spatial concentration \cite{spohn1983long} or thermal \cite{najafi2004forces} gradients. Moving beyond specific models, however, we argue that there are important \emph{generic} features of fluctuation-induced forces that can be fruitfully derived by considering the fluctuation spectrum, and treating it as a phenomenological quantity. 

\section*{Maritime Casimir Effect}

\begin{figure*}
\centering 
\subfigure[]{\includegraphics[width=0.73\columnwidth]{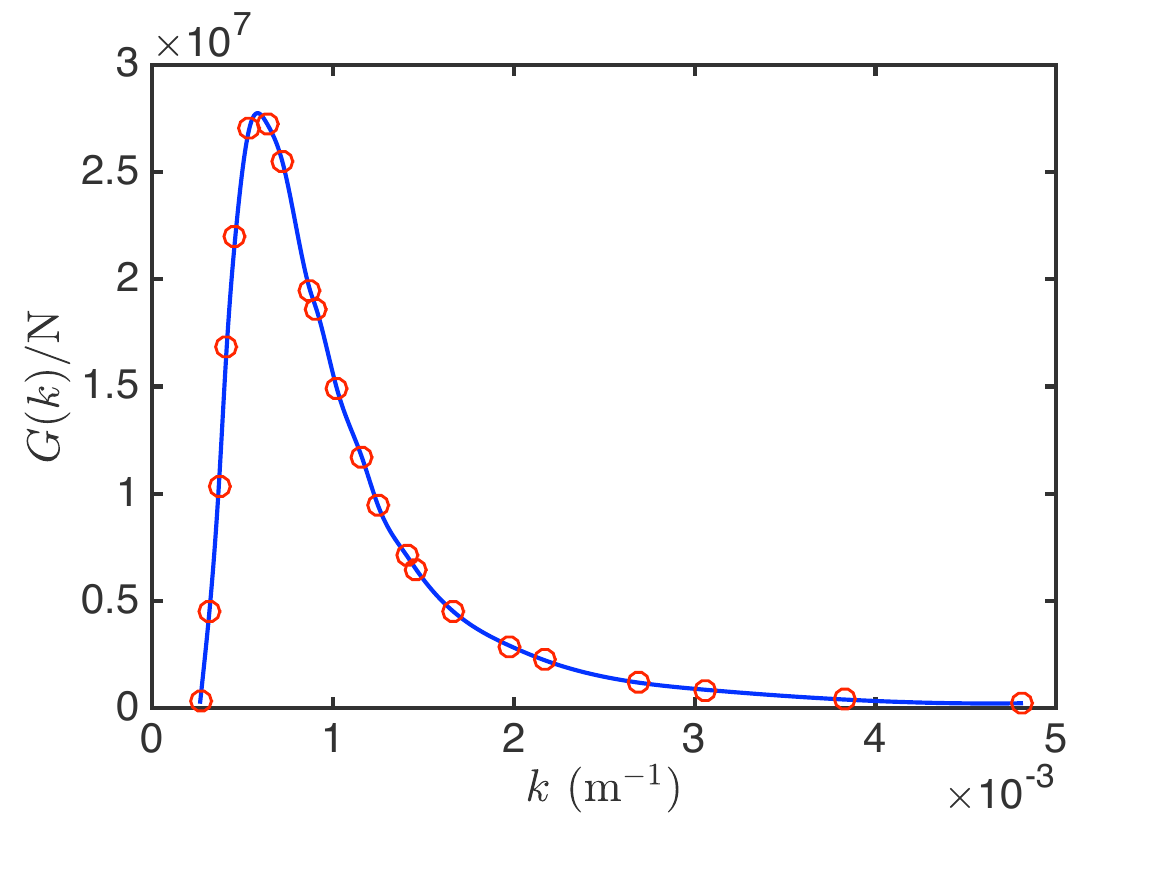}}
\subfigure[]{\includegraphics[width=0.65\columnwidth]{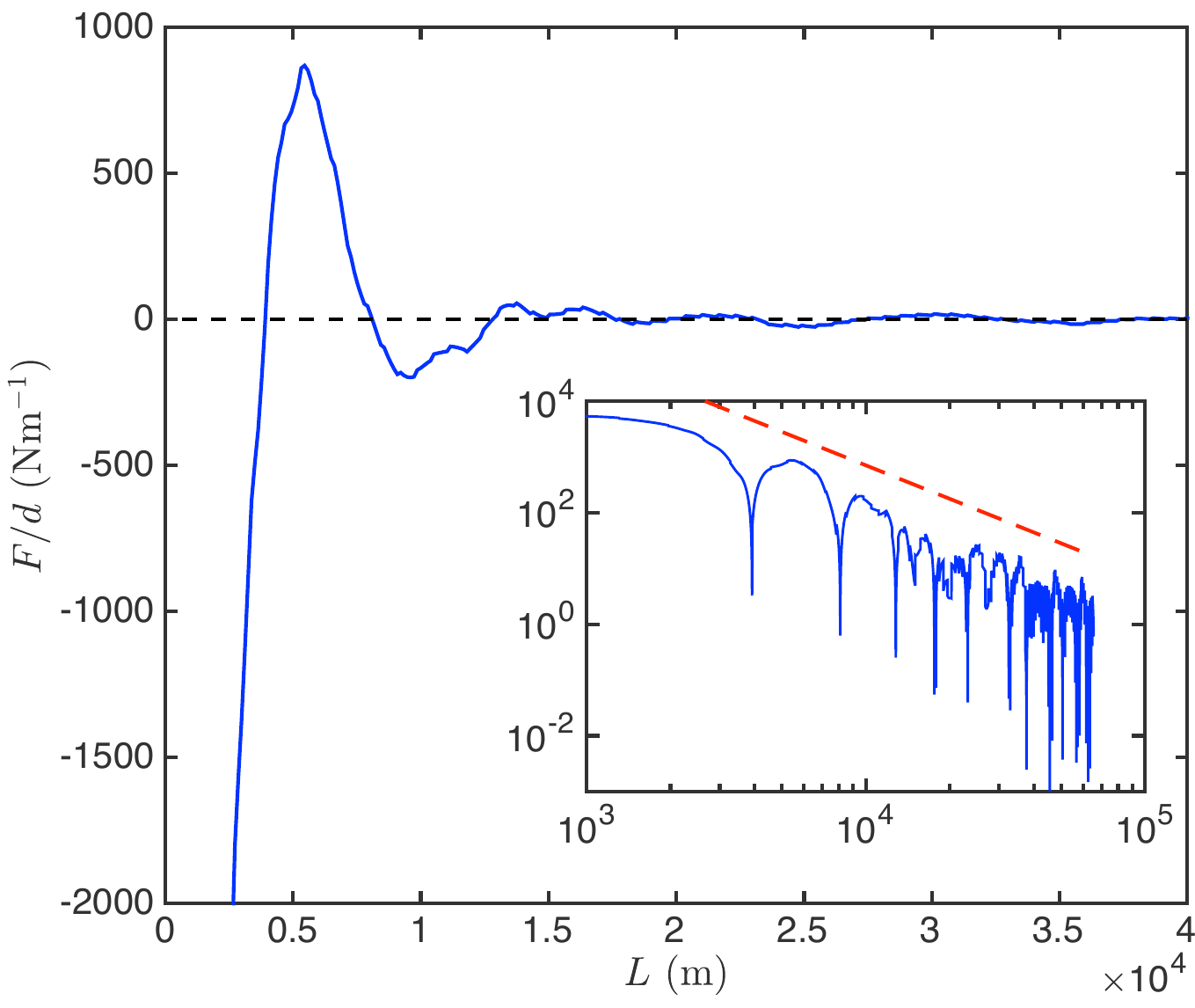}}
\subfigure[]{\includegraphics[width=0.65\columnwidth]{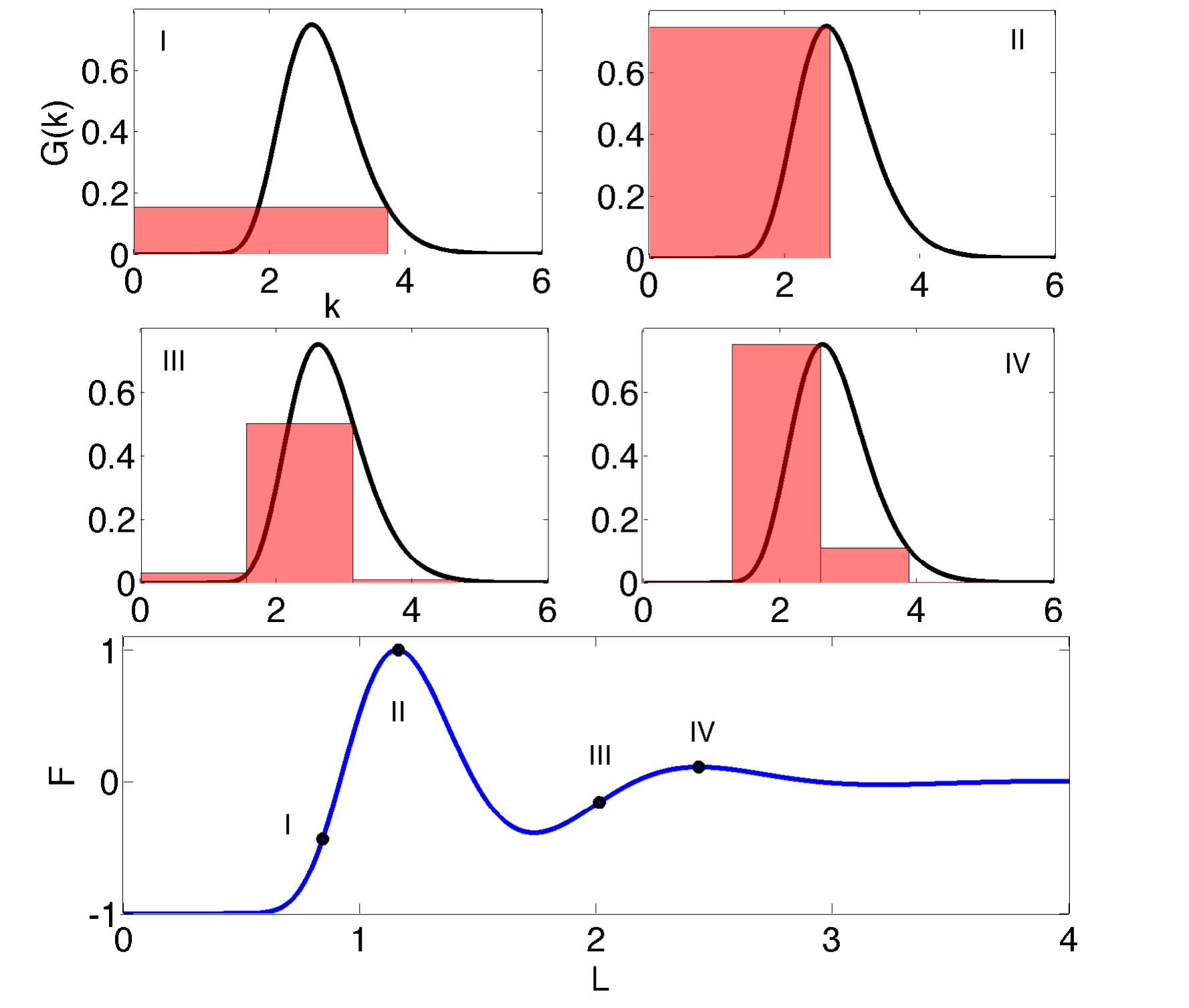}}
\caption{(a) The energy spectrum of ocean waves is non-monotonic. The data is taken from ref.~\cite{pierson1964proposed} for a windspeed of 33.6 knots ($\approx 17 \mathrm{m/s}$) (b) The fluctuation-induced force per unit length. The inset shows that the force (solid blue curve) agrees with the asymptotic prediction Eqn. (\ref{longdist_asym}) (dotted red line). (c) The disjoining force is the difference between the integral over the noise spectrum (area under the curve), and the Riemann sum (the shaded regions); crucially the sum overestimates the integral (i.e.~the force is repulsive) when one ``grid point'' is sufficiently close to the maximum in the distribution, $\kmax\approx n\pi/L$ for some $n$; more often the sum underestimates the integral, leading to attraction. Note that the quantities on the axes are dimensionless.  }
\label{fluct_force}
\end{figure*}

We first illustrate the central result, Eq. (\ref{central_res}), by applying it to the classical  hydrodynamic example of ocean surface waves that are driven to a non-equilibrium steady state via wind-wave interactions. 
We treat the one-dimensional case in which the wind blows in a direction perpendicular to the plates (a simple model of ships on the sea), and hence waves traveling parallel to the plates are negligible. Observations \cite{pierson1964proposed} show that the spectrum $G(k)$ is non-monotonic (see Fig.~\ref{fluct_force}a). While various fits have been proposed \cite{pierson1964proposed,alves2003revisiting}, these are untested at large and small wavenumber. Instead, we compute the force in \eqref{central_res} numerically, approximating the spectrum by a spline through the measured data points of Pierson \& Moskowitz \cite{pierson1964proposed}, and truncating for wavembers beyond their measured ranges.  Figure \ref{fluct_force}(b) shows that the resulting force is non-monotonic and oscillatory as a function of $L$: the force can be \emph{repulsive} ($\Ffluct >0$) as well as attractive ($\Ffluct < 0$). 
Physically, the origin of the attractive force is akin to the Casimir force between metal plates --- the presence of walls restricts the modes allowed in the interior, so that the energy density outside the walls is greater than that inside. This attractive ``Maritime Casimir'' force has been observed since antiquity \citep[see e.g.,][and refs therein]{boersma1996maritime} and experimentally measured in a wavetank \cite{denardo2009water}.  However, the \emph{non-monotonicity} of the spectrum gives rise to an \emph{oscillatory} force-displacement curve. In particular, the force is repulsive when one of the allowed discrete modes is close to the wavenumber at which the peak of the spectral density occurs (see Fig.~\ref{fluct_force}(c)): here the sum overestimates the integral in \eqref{central_res} and the outward force is greater than the inward force. Thus, the local maxima in the repulsive force are approximately located at
\begin{equation}
L_{n} \approx n \frac{\pi}{k_{\mathrm{max}}}, 
\end{equation}
where $G'(k_{\mathrm{max}}) =0$; the separation between the force peaks is $\Delta L \approx \pi/k_{\mathrm{max}}$. In a maritime context, our calculation shows that if the separation between ships is $L>\pi/k_{\mathrm{max}} $, the repulsive fluctuation force will keep the ships away from each other. 

To our knowledge, this is the first prediction of a repulsive Maritime Casimir force, and is yet to be verified experimentally. Clearly quantitative measurement of this oscillatory hydrodynamic fluctuation force in an uncontrolled {\em in-situ} ocean environment influenced by intermittency would be challenging, although the controlled laboratory framework used in pilot-wave hydrodynamics is ideally suited for direct experimental tests \cite[e.g.,][]{bush2015pilot}.   We note that an oscillatory force has been observed in the acoustic analogue for which a non-monotonic fluctuation spectrum was produced \cite{larraza1998acoustic,larraza1998force}. Moreover, one-dimensional filaments in a flowing two-dimensional soap film are observed to oscillate in phase or out of phase depending on their relative separation \cite{zhang2000flexible}, suggesting an oscillatory fluctuation-induced force; visualization of this instability  reveals the presence of waves and coherent fluctuations as the mechanism for force generation, which is the basis of our approach.

We would expect that the fluctuation-induced force vanishes when the fluid is at thermal equilibrium. To test this, we note that a consequence of the equipartition theorem is that the energy spectrum for a three-dimensional isotropic fluid at equilibrium is monotonic, and has the scaling \cite{forster1977large}
\begin{equation}
G_{\mathrm{eq}}(k) \propto k^2. 
\label{g_Eq}
\end{equation}
Noting that in 3D $\delta k = \delta k_x \delta k_y \delta k_z/(4 \pi k^2)$, Eq. (\ref{central_res}) becomes 
\begin{align}
\Ffluct&= \frac{1}{4 \pi} \int_0^{\infty} \mathrm{d}k_y \int_0^{\infty} \mathrm{d}k_z \left( \sum_{m=1}^{\infty}  \; \Delta k - \int_{0}^{\infty}  \; \mathrm{d}k \right) =0, \nonumber \\ 
\label{zero_force}
\end{align}
where we have used the fact that the Riemann sum and integral agree exactly for a constant function. Checking this special case confirms that our approach can, in certain circumstances, distinguish between equilibrium and non-equilibrium: in the continuum hydrodynamic setting, a non-zero fluctuation induced force implies non-equilibrium. We will comment on the ultraviolet divergence (divergence in $G(k)$ as $k \rightarrow \infty$) in Eq. (\ref{g_Eq}) and the breakdown of continuum hydrodynamics in the section below. We note that the result, Eqs. (\ref{g_Eq})-(\ref{zero_force}), applies only to isotropic one component fluids --- thermal Casimir forces exist in systems such as liquid crystals \cite{ajdari1991fluctuation} or liquid mixtures near criticality \cite{fisherphenomena,hertlein2008direct}. 

\section*{General Phenomenology of Narrow Unimodal Spectra}

Importantly, the phenomenology of non-monotonic, and even oscillatory, forces is generic for sufficiently narrow, unimodal spectra. To see this, and to make some general quantitative predictions,  we perform a Taylor expansion of a general uni-modal spectrum, $G(k)$, about its maximum at $k = k_{\mathrm{max}}$, to find that
\begin{equation}
G(k) \approx \begin{cases}
G_0 \left[ 1 - \nu^{-2} (k-k_\mathrm{max})^2 \right], & |k-k_\mathrm{max}|<\nu \\ 
0 & \mathrm{otherwise}, 
\end{cases}
\end{equation}  
where $G_0 = G(k_{\mathrm{max}})$, $G_2 = G''(k_{\mathrm{max}})$ and $\nu = \sqrt{-2G_0/G_2}$ is the peak width based on a parabolic approximation. In the narrow-peak limit ($\nu \ll \pi/L$, $\nu \ll k_{\mathrm{max}}$), the force close to the $n^{th}$ peak is given by
\begin{equation}
F_{n} \approx  \begin{cases}
\frac{G_0 \pi}{4 L} \left[ 1 - \nu^{-2} \left(\frac{n \pi}{L}-k_\mathrm{max} \right)^2 \right] - \frac{ G_0 \nu}{3}, & \left|\frac{n \pi}{L}-k_\mathrm{max} \right|<\nu, \\ 
 - \frac{ G_0 \nu}{3}  & \mathrm{otherwise}. 
\end{cases} 
\label{nthpeak}
\end{equation}  
From the simplified spectrum in Eq. (\ref{nthpeak}) it may be shown  that the $n^{th}$ maximum is located at $L_{n}^{\mathrm{max}} = n \pi/k_\mathrm{max} + O((\nu/k_{\mathrm{max}})^2)$, and has magnitude 
\begin{equation}
F_{n,\mathrm{max}} = \frac{G_0 \pi}{4 L} - \frac{ G_0 \nu }{3} = \frac{G_0 k_\mathrm{max}}{4 n} - \frac{ G_0 \nu}{3} .
\label{max_asym_force}
\end{equation} 
Thus, the maximum force is linear in inverse plate separation and the force  reaches its minimum when
\begin{equation}
k_{\mathrm{max}} - \frac{n \pi}{ L} = \nu. 
\end{equation}
Writing  $L = L_{n}^{\mathrm{max}} +l_n =  n \pi/k_\mathrm{max} + l_n$, where $l_n$ is the half-width of the peak in force, we obtain
\begin{equation}
l_n  = n \pi \left(  \frac{1}{k_\mathrm{max}} - \frac{1}{\nu + k_\mathrm{max}} \right)\approx \frac{n \pi\nu}{\kmax^2}.
\label{peak_width}
\end{equation}
Therefore the width of the force maxima increases \emph{linearly} with $n$, and the positions of the $n^{th}$ mechanical equilibria ($\Ffluct =0$) in the limit of narrowly-peaked spectra ($\nu \ll k_{\mathrm{max}}$) are given by
\begin{equation}
L_{n,eq} \approx L_n \pm l_n\approx n \pi \left( \frac{1}{k_{\mathrm{max}}} \pm \frac{\nu}{\kmax^2}\right).
\label{sta_unsta_eq_eq}
\end{equation}
Here the positive (negative) branches correspond to stable (unstable) equilibria. Eqs. (\ref{max_asym_force}) and (\ref{peak_width}) predict that the force-displacement curve has peak repulsion $\propto 1/L$ and peak width $\propto n \propto L$ for $L\ll \pi/\nu$.

The asymptotic prediction \eqref{max_asym_force} arises from assuming that only one term in the sum \eqref{eqn:Fin} is significant. As such, this approximation breaks down when the width of the rectangles becomes comparable to the width of the peak in $G(k)$ itself, i.e.~when $L\sim1/\nu$. In consequence, the prediction of \eqref{max_asym_force} that the force  becomes monotonically negative for $L> L_{\mathrm{thres}} = 3 \pi / 4 \nu$ is incorrect. However, in the limit $L \gg \pi/\nu $, the Riemann sum in (\ref{central_res}) is non-zero only for $L (k_{\mathrm{max}} - \nu)/\pi \lessapprox m \lessapprox L (k_{\mathrm{max}} + \nu)/\pi$; the force then continues to oscillate between attractive and repulsive and has the asymptotic decay
\begin{equation}
F_{\mathrm{min}} \sim - \frac{\pi^2 G_0}{3 \nu} \frac{1}{L^2}, 
\label{longdist_asym}
\end{equation} 
which is the minimum (or maximal attractive) force.  The inverse square decay is shown in the inset of Fig.~\ref{fluct_force}b.

These predictions are borne out by the numerical results for the Maritime Casimir effect discussed earlier (see fig.~\ref{fluct_force}b), but more importantly form a phenomenological theory that can be applied to systems where the fluctuation spectrum is not known \emph{a priori}: if force measurements are found to illustrate these scalings then we suggest that the underlying spectrum is likely to be narrow and uni-modal \footnote{The scalings derived here are specialized to the case of interactions between plates, which is a reasonable approximation for the interaction between objects when their separation is much less than their radii of curvature. }.

We can now revisit the case of classic fluids at equilibrium. Obviously, the divergence in Eq. (\ref{g_Eq}) as $k\rightarrow \infty$ is unphysical. This ultraviolet divergence is cured by noting that hydrodynamic fluctuations, as captured by the spectrum $G(k)$, are suppressed at the molecular lengthscale $k  \sim 2\pi/\sigma$ where $\sigma$ is the molecular diameter. Therefore, our analysis (Eq. (\ref{nthpeak})) predicts an oscillatory fluctuation-induced force with a period that is comparable to the molecular diameter. This is indeed observed in confined equilibrium fluids  \cite{horn1981direct}, although clearly at the molecular lengthscale our hydrodynamic description breaks down and other physical phenomena, such as proximity induded layering, become relevant. Importantly, while the oscillation wavelength of the disjoining force in equilibrium fluids is nanoscopic, of order the molecular scale, the oscillation wavelength in active non-equilibrium systems can be much larger than the size of the active particle, because the mechanism of force generation lies in a non trivial partition of energy. 

\section*{Force Generation with Active Brownian Particles}

\begin{figure*}
\centering
\subfigure[]{\includegraphics[width=0.68\columnwidth]{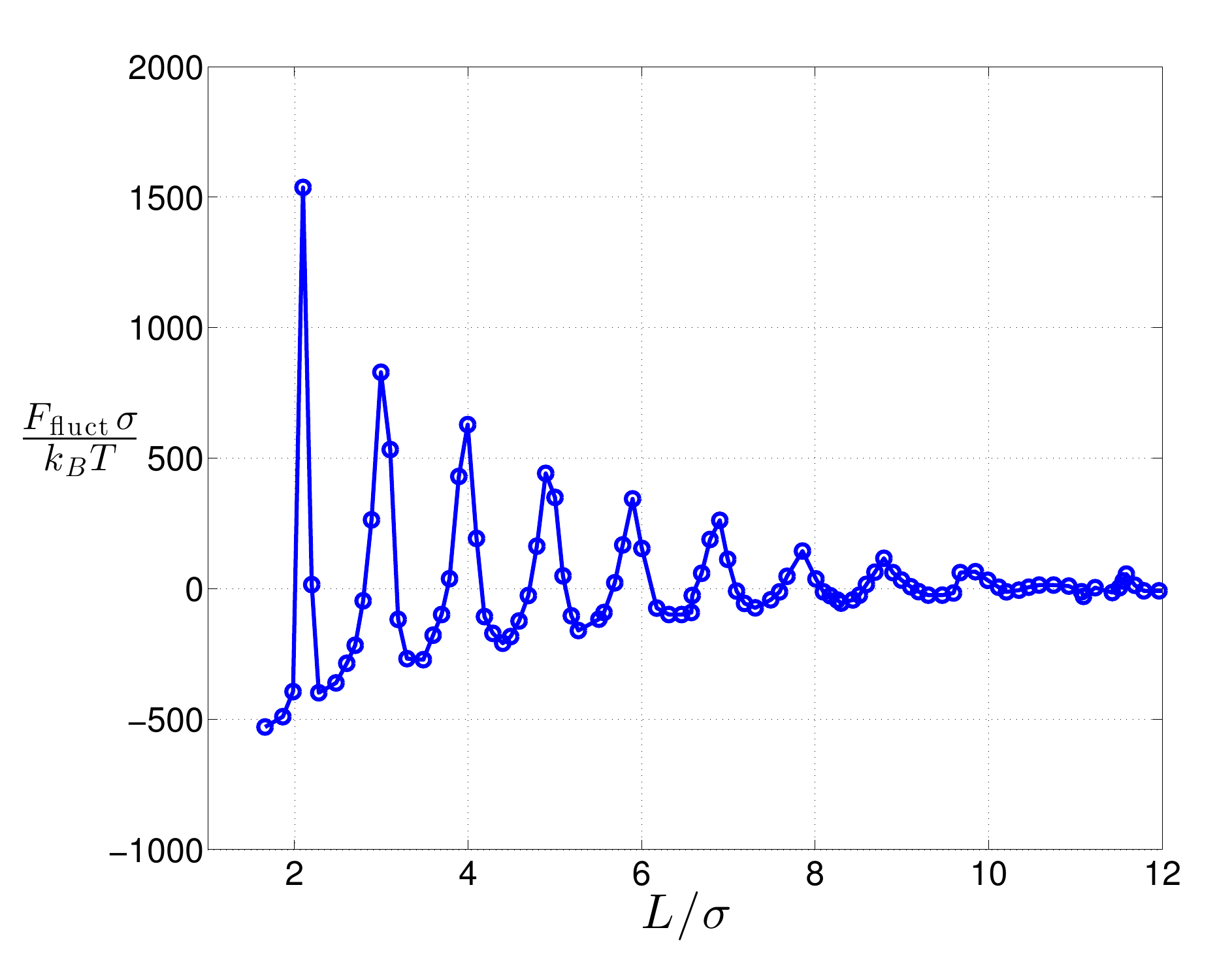}}
\subfigure[]{\includegraphics[width=0.68\columnwidth]{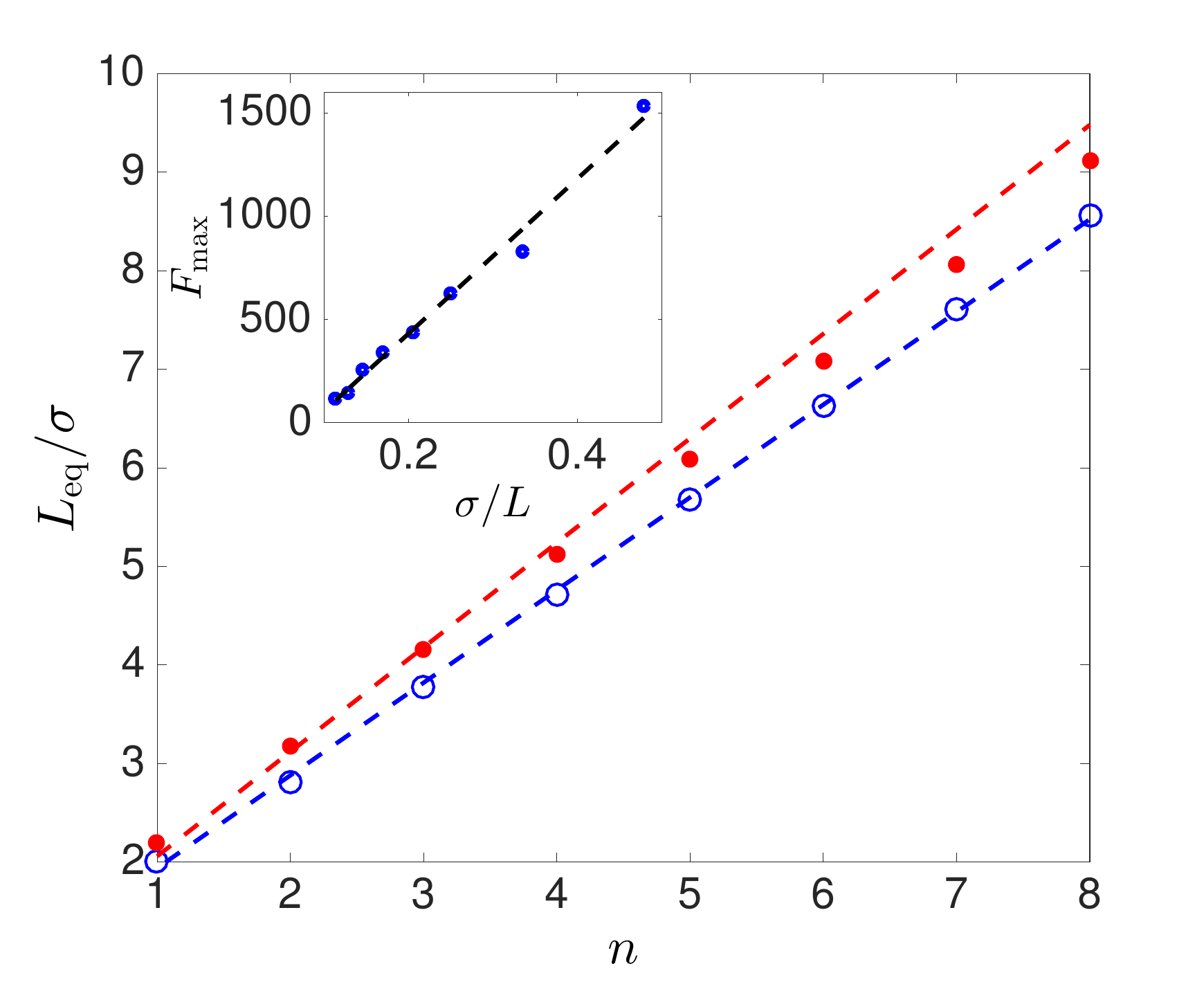}}
\subfigure[]{\includegraphics[width=0.68\columnwidth]{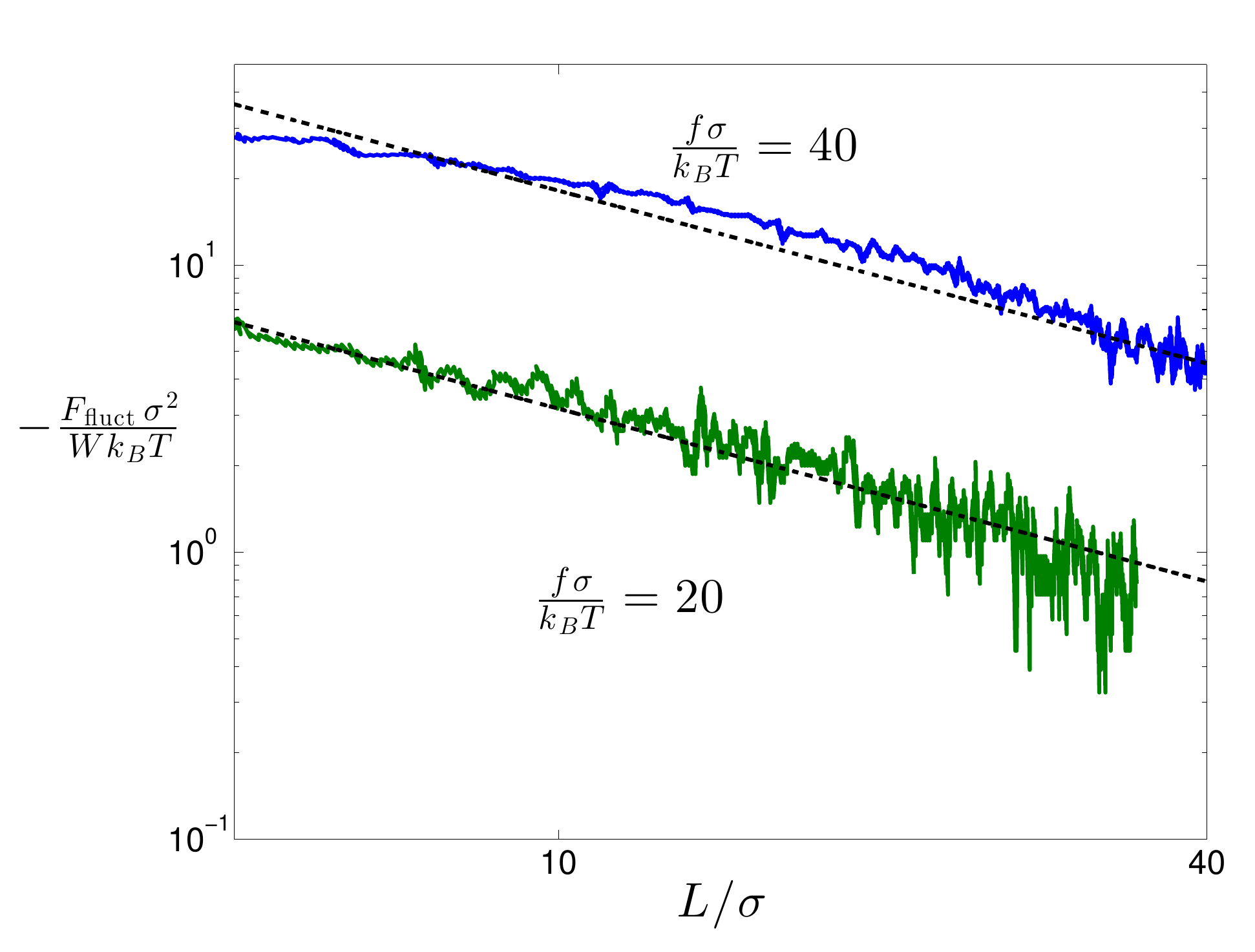}}
\vspace{-0.45cm}
\caption{ Comparison of our theory with the simulations of a 2D suspension of self-propelled Brownian spheres, confined between hard slabs, that interact via the Weeks-Chandler-Anderson potential \cite{ni2015tunable}. In (a) and (b) the packing fraction in the bulk is $\rho \sigma^2 = 0.4$, where $\sigma$ is the particle diameter, the wall length is $W=10 \sigma$, and self-propulsion force  $f = 40 k_B T/\sigma$. (a) The raw force-displacement curve for $\rho \sigma^2 = 0.4$ from \cite{ni2015tunable}.  (b) When replotted as suggested by our asymptotic predictions (\ref{max_asym_force}) and (\ref{peak_width}) these data suggest that the underlying fluctuation spectrum is unimodal and has a narrow peak, with parameters $G_0 = 4.8 \times 10^{3}$ and $\nu = 0.2/\sigma$. (As the peaks are spaced approximately $\sigma$ apart, we assume $k_{\mathrm{max}} = \pi/\sigma$, and $G_0$ and $\nu$ are obtained from fits of \eqref{sta_unsta_eq_eq} to the simulation data.) The positions of the stable (closed circles) and unstable (open circles) mechanical equilibria (when $\Ffluct= 0$) are given by ${L}_{\mathrm{eq}}$, and the dotted lines are theoretical predictions (Eq. (\ref{sta_unsta_eq_eq})). The inset shows the force maxima in (a) $\propto 1/L$ and agrees with Eq. (\ref{max_asym_force}). (c) For ideal non-interacting self-propelled point particles, the function $A \sigma/L$ (black dotted line, \emph{c.f.,}  Eq. (\ref{f_fluct_ideal})) can be fitted (using $A$) to simulation data with $F \sigma^2/(W k_B T) = 40$ ($A = 182$) and $F \sigma^2/(W k_B T) = 20$ ($A = 31.6$). Here $W = 80 \sigma$.} 
\label{abpa}
\end{figure*}

Interestingly, our asymptotic results are in agreement with force generation in what one might consider to be the unrelated context of self-propelled active Brownian particles. Ni \emph{et al.} \cite{ni2015tunable} simulated self-propelled Brownian hard spheres confined between hard walls of length $W$  and found an oscillatory decay in the disjoining force (Fig.~\ref{abpa}a). Although this system is two-dimensional, our analysis can be generalized: In 2D, $\delta k = \delta k_x \delta k_y/(2\pi k)$, and hence 
\begin{equation}
F_{\mathrm{in}} = \frac{1}{4} \sum_{n=1}^{\infty} \Delta k \int_{0}^{\infty}   \frac{G\left(\sqrt{(n \Delta k)^2 + q^2} \right)}{2 \pi \sqrt{(n \Delta k)^2 + q^2}} \mathrm{d} q.
\end{equation}
However, we can redefine 
\begin{equation} 
h(k) \equiv \int_{0}^{\infty}  \frac{G(\sqrt{q^2 + k^2})}{2 \pi \sqrt{q^2 + k^2}} \; \mathrm{d}q
\end{equation}
as an effective 1D spectrum and substitute $h(k)$ for $G(k)$ in Eq. (\ref{central_res}). Performing the same asymptotic analysis as for the narrow-peak limit, the asymptotic scalings (\ref{max_asym_force}) and (\ref{peak_width}) are reproduced, in quantitative agreement with simulations. (we note that the linear scaling shown in Fig.~\ref{abpa}b also implies that the width of the peak scales linearly in $L$, as predicted by Eq.~(\ref{peak_width})). The $\sim 1/L^{2}$ decay expected for large $L$ is not observed in the data as the asymptotic approximations underlying Eq.~(\ref{max_asym_force}) only break down for $L\gtrsim L_{\mathrm{thres}} \approx 12 \sigma$, with $\sigma \nu = 0.2$ estimated from the data.  This agreement between the data and our asymptotic framework suggests that the underlying spectrum for active Brownian systems is narrow and non-monotonic\footnote{For smaller values of the active self-propulsion force $f$ simulated in \cite{ni2015tunable}, the peaks are less pronounced and obscured by numerical noise.}. The slight discrepancy with the linear fit at large $n$ is likely due to the fact that our asymptotic scaling only holds in the regime $L \ll \pi/\nu$ (note that $\pi/\nu \approx 15 \sigma$ in Fig.~\ref{abpa}b, and the linear fit deteriorates when $L\gtrsim7\sigma$, confirming that the value of $\nu$ estimated from fitting to the  width and height of the force peak is at least of the correct order of magnitude). An additional source of the discrepancy may be that the signal-to-noise ratio decreases for increasing plate separation as the magnitude of the force becomes smaller. 

Further analytical insights can be obtained by considering the limit of no excluded volume interaction between particles in which Ni \emph{et al.}  \cite{ni2015tunable} observed that the disjoining pressure is attractive and decays monotonically with separation (similar results have been obtained by Ray \emph{et al.} \cite{ray2014casimir} for run-and-tumble active matter particles). 
This observation can be explained within our framework by noting that the self-propulsion of point-particles induces a Gaussian colored noise  $\zeta(t)$ satisfying \cite{farage2015effective}
\begin{equation}
\left< \zeta(t)\right> = 0, \; \; \left< \zeta(t)\zeta(t')\right> = \frac{f^2}{3} e^{-2 D_r |t-t'|}, 
\end{equation} 
where $f$ is the active self-propulsion force and $D_r$ is the rotational diffusion coefficient. In the frequency domain, the fluctuation spectrum $S(\omega)$ is the Fourier transform of the time-correlation function and is
\begin{equation}
S(\omega) = \frac{4 D_r f^2}{3} \frac{1}{4 D_r^2 + \omega^2}.  
\label{S_colour}
\end{equation}
The Lorentzian noise spectrum of Eq. (\ref{S_colour}) deviates from entropy-maximizing white noise. Assuming a linear dispersion relationship, $\omega \propto k$, we note that because the spectrum of Eq. (\ref{S_colour}) is now monotonic, the difference between the integral and the Riemann sum, Eq.~\ref{central_res}, is monotonic and $\sim 1/L$.  Now, the degree of freedom in the direction parallel to the plates can be integrated yielding 
\begin{equation}
\Ffluct \propto -\frac{f^2}{L}, 
\label{f_fluct_ideal}
\end{equation}
for large $L$. Hence, we expect to see a monotonic force--displacement relation, as observed by Ni \emph{et al.} \cite{ni2015tunable}.
Indeed, Fig.~\ref{abpa}(c) shows that the disjoining pressure obtained from simulations is consistent with this scaling: the decay $\propto 1/L$ and doubling the activity $f$ increases the prefactor by a factor of $5.6$, very nearly the predicted factor of $4$.  (We believe the slight discrepancy to be caused by the sampling noise, which as seen in Fig.~\ref{abpa}(c) is  especially significant at large plate separations and is sufficient to alter the estimate of the fitting parameter.) Since oscillatory force decay is only seen for finite, active particles, evidently  the coupling between excluded volume interactions and active self-propulsion gives rise to the non-monotonic spectrum and the oscillatory decay seen in Fig.~\ref{abpa}(a). In particular, the presence of excluded volume interactions give rise to a length scale of energy injection -- the particle diameter -- and indeed the peak in the spectrum, $k_{\mathrm{max}}$, is approximately the inverse particle diameter. 

Non-monotonic energy spectra are also found in the continuum hydrodynamic description of active particles \cite{wensink2012meso,bratanov2015new}, as well as active swimmers in a fluid \cite{slomka2015generalized}. For a wide class of such ``active turbulent'' systems, the fluctuation spectra take the analytical form \cite{bratanov2015new}
\begin{equation}
G(k) = E_0 k^{\alpha} e^{- \beta k^2 },
\label{bratanov}
\end{equation}
where $E_0$, $\alpha$ and $\beta $ are constants that depend on the underlying microscopic model. This spectrum is narrowly peaked when $\alpha/\beta\gg 1/\beta$, i.e. $\alpha\gg1$. Although Eq. (\ref{bratanov}) captures the fluctuations of the active species, but not the background fluid, numerical results show that the energy spectrum of the background fluid -- the spectrum that enters into our framework -- is {\em also} non-monotonic \cite{slomka2015generalized}. Therefore, our asymptotic framework, Eqs. (\ref{max_asym_force}) -- (\ref{sta_unsta_eq_eq}), derived for a general unimodal spectrum, can also be applied to those systems. We note that the effective viscosity of an active fluid in a plane-Couette geometry has been shown numerically \cite{slomka2016geometry} to be an oscillatory function of plate separation; this supports the oscillatory force framework reported here. Furthermore, oscillatory and long-range fluctuation-induced forces have been reported in other soft matter systems, including inclusions in a shaken granular medium \cite{cattuto2006fluctuation,lietor2017casimir} (where the density field of the granular medium is also directly shown to be inhomogeneous and oscillatory, qualitatively agreeing with our fluctuating modes framework) and rotating active particles on a monolayer \cite{aragones2016elasticity}. Experimental or numerical measurements of Casimir forces in active systems will serve as a test-bed of our formalism, while the application of our fluctuation spectrum approach to  the phenomenology of other systems are the subject of future work. 

\section*{Conclusion}

There are of course a plethora of ways to prepare non-equilibrium systems.  We suggest that an organizing principle for force generation is the fluctuation spectrum --- the active species drives a non-equipartition of energy. By adopting this top-down view, we computed the relationship between the disjoining pressure and the fluctuation spectrum, and verified our approach by considering two seemingly disparate non-equilibrium physical systems: the Maritime Casimir effect, which is driven by wind-water interactions, and the forces generated by confined active Brownian particles. Our framework affords crucial insight into the phenomenology of both driven and active non-equilibrium systems by providing the bridge between microscopic calculations \cite{takatori2014swim,solon2015pressure,yan2015swim}, measurements of the fluctuation spectra \cite{chu1974laser} and the varied measurements of Casimir interactions \cite{lamoreaux1997demonstration, munday2009measured, sushkov2011observation}.   Although this article is motivated by biological and biomimetic settings, measurements of the non-equilibrium electromagnetic Casimir effect, such as the force that an (active) oscillating charge exerts on a neighboring charge, may also test our theory.

In particular, while the fluctuation spectrum of equilibrium fluids vanishes at the molecular scale, so that force oscillations are seen at the molecular lengthscale (e.g. \cite{horn1981direct}), it is the case that a hydrodynamic system with a force oscillation wavelength much larger than the molecular lengthscale {\em must be out of equilibrium} (because the thermal fluctuation spectrum, $G\sim k^2$, is monotonic). As a corollary, out-of-equilibrium systems can exhibit force oscillations with wavelengths significantly longer than the size of the active particles. More generally, because time reversal symmetry requires equilibrium \cite{pomeau1982symetrie}, it would appear prudent to examine the time correlations in the systems we have studied here. Additionally,  another form of an ``active fluid'' can be constructed in a pure system using, for example, a thermally non-equilibrium steady state; temperature fluctuations in such a system have been observed to give rise to long-range Casimir-like behavior \cite{kirkpatrick2013giant,aminov2015fluctuation}. Hence, an intriguing possibility suggested by our analysis is that rather than tuning forces by controlling the nature (e.g., dielectric properties \cite{french2010long}) of the bounding walls, one can envisage actively controlling the fluctuation spectra of the intervening material.  Indeed, a natural speculation is that swimmers in biological (engineering) settings could (be designed to) actively control the forces they experience in confined geometries.

\acknowledgments
This work was supported by an EPSRC Research Studentship, Fulbright Scholarship and George F. Carrier Fellowship (AAL) and by the European Research Council (Starting Grant GADGET No.~637334 to DV). JSW acknowledges support from Swedish Research Council Grant No. 638-2013-9243, a Royal Society Wolfson Research Merit Award, and the 2015 Geophysical Fluid Dynamics Summer Study Program at the Woods Hole Oceanographic Institution (National Science Foundation and the Office of Naval Research under OCE-1332750).


\end{document}